\documentclass[runningheads]{llncs}

\usepackage[T1]{fontenc}
\usepackage{graphicx}
\usepackage{xurl}
\usepackage[breaklinks=true]{hyperref}

\begin{document}

\title{Reproducibility is the New Copyleft:\\
  Defining AGI-oriented Reproducible Builds}

\titlerunning{Reproducibility is the New Copyleft}

\author{Masayuki Hatta\inst{1}\orcidID{0000-0002-6700-7341}}

\authorrunning{M. Hatta}

\institute{Surugadai University, Hanno Saitama 3570046, Japan\\
\email{hatta.masayuki@surugadai.ac.jp}}

\maketitle

\begin{abstract}
The concept of \emph{copyleft}, as implemented in licenses such as the GNU General Public License, was a legal hack that used copyright to guarantee user freedom by tying the availability of source code to every act of distribution. Its normative force rested on an implicit technical premise: that source code and object code stand in a well-defined, humanly auditable, and reproducible relationship. Large language models and, prospectively, Artificial General Intelligence (AGI) systems systematically violate this premise. The training artifacts jointly required to reconstruct a given model---code, data, weights, hyperparameters, toolchain, and hardware configuration---are each subject to independent legal, technical, and economic constraints that no current open-source framework fully resolves. Sufficiently capable AI systems can also \emph{rewrite} licensed source into functionally equivalent derivatives stripped of their original obligations, a form of laundering against which copyleft has no effective defense.

This paper argues that a functional analogue of copyleft for AGI must be grounded not in share-alike clauses over code, but in \emph{reproducible builds}: a practice guaranteeing bit-exact reconstructability from declared inputs. We review the history and logic of copyleft, critically examine Maffulli's \emph{Second Liberation} thesis according to which AI fulfills Stallman's dream, and show that the argument collapses unless AGI systems are themselves reproducible. Drawing on the Open Source AI Definition (OSAID), the Model Openness Framework (MOF), OpenMDW, and deterministic-inference research by Thinking Machines Lab, SGLang, and others, we define seven requirements for \emph{AGI-oriented reproducible builds}. We further argue that the Model Context Protocol (MCP) and analogous AI-to-AI coupling mechanisms constitute a new \emph{dynamic linking layer} for which copyleft-style licensing is ill-suited, and that Masnick's ``protocols, not platforms'' framework offers a more promising governance template for the AI linking layer.

\end{abstract}

\section{Introduction}

The governance of advanced artificial intelligence has become one of the defining policy questions of the mid-2020s. Within that discussion, two ostensibly distinct strands have evolved largely in parallel: the \emph{open source AI} debate, concerned with the meaning of ``open'' for machine learning systems that combine code, weights, and data; and the \emph{AGI safety and governance} debate, concerned with systems capable of recursive self-improvement or self-replication \cite{hatta2025agi}. This paper argues that these strands intersect at a single, surprisingly technical point: the question of what it means for an AI system to be \emph{reproducible}, and whether reproducibility can play the role for AGI that copyleft played for traditional software.

The original copyleft hack, engineered by Richard Stallman in the 1980s, used copyright law to guarantee that every recipient of a GNU program would also receive the source code necessary to study and modify it \cite{stallman2002free}. As Maffulli has recently argued \cite{maffulli2026second}, copyleft was in this sense a ``permission slip'' for freedom rather than its substantive guarantee: the user still needed the skill and the time to exercise the freedoms the license protected. Maffulli's provocation is that modern AI coding assistants are the real, technical implementation of software freedom---the ``second liberation''---because they collapse the knowledge barrier that made the GPL a merely notional promise for non-programmers.

The argument is rhetorically powerful but substantively incomplete. Maffulli himself concedes, in a postscript, that it would be ``a lot better if the AI code generators were Open Source AI.'' Everything hangs on that ``open'' qualifier. If the tools supposed to realize software freedom are themselves opaque black boxes run on remote servers, the move has merely replaced one form of vendor tyranny with another.

In an earlier essay \cite{hatta2024copyleft}, we suggested that the essence of copyleft was not the share-alike clause per se but the \emph{equivalence} it induced between source and object code: because the source had to accompany the binary as the ``preferred form for modification,'' the binary was in principle always reducible to a human-readable, collaboratively maintainable representation. Generative AI breaks this equivalence. Even when training code is fully released, the weights, training data, preprocessing scripts, random seeds, GPU microcode, and distributed-scheduling decisions jointly determine the resulting model, and no subset of these alone suffices to reconstruct it. The natural technical analogue of copyleft in this setting is the \emph{reproducible builds} methodology: a discipline guaranteeing that a specified set of inputs always produces a bit-identical output, so that any third party can verify the correspondence between a published artifact and its stated sources.

This paper develops that proposal into a set of concrete requirements for AGI systems, but also extends the diagnosis in two directions. First, generative AI creates a new and corrosive problem for copyleft itself: sufficiently capable assistants can \emph{rewrite} GPL-licensed source into functionally equivalent code releasable under any license, collapsing the cost asymmetry on which copyleft's practical force has always depended. Second, the rapid adoption of the Model Context Protocol (MCP) and similar runtime coupling standards establishes a new \emph{dynamic linking layer} between AI systems, for which the appropriate governance framework is not copyleft at all but the ``protocols, not platforms'' template developed by Masnick in the social-media context. The paper thus proposes a \emph{two-layer governance architecture}: reproducible builds for the production layer and protocol governance for the linking layer. Making the underlying governance logic of this architecture explicit---a gap that OSAID, MOF, the EU AI Act's open-source provisions, and the Reproducible Builds project share---is the primary contribution of this paper.

\section{Copyleft: Definition, History, and Underlying Logic}
\label{sec:copyleft}

Copyleft is commonly defined as a licensing technique that uses the exclusionary power of copyright in reverse: rather than restricting copying, the license permits copying and modification on the condition that derivative works be distributed under identical terms \cite{stallman2002free}. The GNU General Public License (GPL), first released in 1989 and revised as GPLv3 in 2007, is the canonical implementation \cite{fsf2007gpl3}. The Free Software Definition codifies four freedoms: to run, to study and modify, to redistribute, and to redistribute modified versions \cite{fsf2024freedoms}. Crucially, the freedoms to study and to redistribute modified versions (freedoms~1 and~3 in the FSF's zero-indexed numbering) are conditioned on ``access to the source code'' as a \emph{precondition}; the GPL's share-alike obligation ensures that this precondition propagates along every distribution chain.

What is less often emphasized is the technical premise that makes the Free Software Definition intelligible in the first place. The definition presupposes that \emph{source code} is a well-defined artifact standing in a specific relation to the \emph{object code} that ends up executing on a user's machine. GPLv3 makes this explicit by defining the ``Corresponding Source'' as ``all the source code needed to generate, install, and \ldots\ run the object code and to modify the work,'' including scripts to control compilation and installation \cite{fsf2007gpl3}. The license assumes, in other words, that a deterministic, human-auditable build process converts source into binary, and that obligations attached to the source therefore meaningfully constrain what users can do with the binary.

This is the technical foundation on which the legal structure rests. Copyleft enforces the equivalence of source and object: whatever the user possesses at runtime must be derivable, via a disclosed and reproducible procedure, from the disclosed source. When that equivalence holds, the four freedoms are genuinely actionable; when it fails, the license may still be formally satisfied while the substantive freedoms evaporate. It is this equivalence, rather than share-alike as such, that deserves to be regarded as the normative core of copyleft---and that the present paper takes as its starting point.

\section{The Collapse of Source--Object Equivalence in Modern AI}
\label{sec:collapse}

Machine learning systems---and above all large language models---systematically violate the assumption that object code is the deterministic output of a disclosed source-code compilation process. A trained model's observable behavior is jointly determined by at least the following:

\begin{enumerate}
    \item the training code, including data-loading, tokenization, optimizer, and loss-computation logic;
    \item the training dataset, often composed of billions of documents from heterogeneous sources with varying legal status;
    \item the set of hyperparameters, including learning rates, batch sizes, dropout probabilities, and seed values;
    \item the sequence of random numbers actually drawn during training, which depends not only on seeds but on the scheduling behavior of distributed workers;
    \item the software toolchain, from the deep-learning framework down to CUDA/cuDNN, compiler versions, and floating-point intrinsics;
    \item the hardware, including the specific GPU microarchitecture, driver, and interconnect topology;
    \item the model weights produced by the above, typically numbering in the billions and stored in formats whose numerical semantics depend on item~5.
\end{enumerate}

\noindent A system that releases only the weights, or only the training code, or only the dataset, does not enable a third party to reconstruct the model. It does not even enable them to verify whether the released weights were in fact produced by the disclosed training procedure. The four freedoms, in the Free Software sense, become unexercisable: one cannot meaningfully study the system's behavior, modify it in a principled way, or redistribute a known-good version, if none of the claimed inputs can be independently checked.

The Open Source AI Definition (OSAID) promulgated by the Open Source Initiative in October 2024 \cite{osaid2024} explicitly recognizes this problem. The OSAID requires that an open source AI system make available: (i) the complete source code used to train and run the system; (ii) the model parameters, including weights; and (iii) sufficiently detailed \emph{data information} to allow a ``skilled person'' to build a substantially equivalent system using the same or similar data. The data-information category is the OSAID's pragmatic compromise with the legal fact that many training corpora cannot be redistributed, whether because they contain personally identifiable information, because they were scraped under text-and-data-mining exceptions that do not extend to redistribution, or because they incorporate copyrighted material \cite{osaid_faq,legalblogs2025osaid}. The OSI has continued to refine OSAID and made data governance its central priority for 2025--2026 \cite{osi2025datagov,osi2025report}.

This compromise has been controversial. Critics such as the Software Freedom Conservancy have argued that ``data information'' falls short of the ``preferred form for modification'' standard that the original Open Source Definition demands, because a substantially equivalent dataset is not the same as the original and the OSAID therefore fails to require reproducibility of the scientific process by which the system was built \cite{kuhn2024osaid}. Parts of the Debian community have taken the further view that most models blessed as OSAID-compliant would be ineligible for distribution under the Debian Free Software Guidelines, because their training data is not redistributable as ``source'' \cite{kuhn2024osaid}. The Linux Foundation's Model Openness Framework (MOF) \cite{white2024mof} and the associated Model Openness Tool \cite{mot2025} take a different approach, defining three tiers---Open Model, Open Tooling, and Open Science (Class~I being the most complete)---with the highest tier demanding raw training datasets, intermediate checkpoints, and log files. OpenMDW, a license issued alongside MOF, is notably \emph{permissive rather than copyleft} \cite{linuxfoundation2025openmdw}---a design decision that this paper's argument suggests may need to be revisited for AGI.

What is interesting, for present purposes, is that none of these frameworks guarantees the source--object equivalence that copyleft presupposed for software. Even a Class~I (Open Science) MOF release, with all 17~lifecycle components disclosed, cannot by itself ensure that a user rebuilding the model from those components will obtain bit-identical weights. The gap between ``all ingredients disclosed'' and ``system verifiably rebuildable'' is precisely where the copyleft analogy has to be reconstructed on a new technical foundation.

\section{Maffulli's Second Liberation Thesis and Its Limits}
\label{sec:maffulli}

Maffulli \cite{maffulli2026second} advances a striking thesis. The GPL, on his reading, was a legal hack that protected users from vendor tyranny but left them dependent on a scarce supply of expert developers to exercise that protection. AI code assistants now democratize that expertise. When a non-programmer can instruct an agent to refactor an abandoned library in an afternoon, the right to fork ceases to be theoretical. Borrowing a framing from Armin Ronacher that Maffulli approvingly cites, software is shifting from a \emph{static monument} maintained by an elite to a \emph{fluid resource} reshapable by any user. In this sense, AI provides the ``technical enforcement'' of the freedoms that copyleft could protect only legally.

There is much to this. The distinction between formal and substantive freedom is real, and the reduction of the skill premium required to modify code is plainly happening, at least at the margin. Yet the thesis is incomplete in a way that Maffulli himself half-acknowledges: the argument is sound only if the AI tools doing the liberation are themselves open source AI. If the code assistant is a remote API controlled by a single vendor, the user who has formally gained the freedom to fork has in fact acquired a new dependency---the vendor can change the model, withdraw the service, censor refactorings, log codebases, or silently inject behavior the user cannot audit. Widder, West, and Whittaker \cite{widder2023open} argue that the political economy of ``open'' cloud AI systematically favors concentrated incumbents, and these runtime risks are among its concrete expressions. The substantive freedom the Second Liberation promises is, in such a setting, merely displaced from the maintainer of the upstream library to the operator of the assistant.

A second limitation is epistemic. Even if the model is open-weight and its training code released, a user who relies on the model's output to reshape software has no way to verify that the model's behavior was not shaped, during training, by adversarial data or supply-chain attacks. Both training-data extraction \cite{carlini2021extracting} and training-data poisoning \cite{carlini2023poisoning} are well-documented attack vectors, and models trained on scraped web-scale corpora offer a particularly large surface for the latter. A user cannot audit a trillion-parameter model by reading it. The only available form of audit is procedural: verify that the claimed inputs produce the claimed outputs. This is exactly the role that copyleft-style source availability once played for ordinary software. The fix for Maffulli's diagnosed gap is not simply ``more powerful AI,'' but AI whose construction is as transparently reconstructible as a GPL-licensed compiler.

\section{The Rewrite Problem: AI as a Copyleft-Laundering Machine}
\label{sec:rewrite}

There is a further, and more corrosive, implication of Maffulli's Second Liberation that Maffulli himself does not draw out. If AI coding assistants can refactor arbitrary codebases on demand, then the same capability that liberates the user from the tyranny of the maintainer also permits the systematic laundering of copyleft obligations. A user who possesses a GPL-licensed library and wishes to incorporate its functionality into a proprietary product need no longer negotiate with upstream, nor reason about the boundary between derivative work and aggregation. They can instruct an AI assistant to produce a functionally equivalent rewrite---clean-room by construction, or at least plausibly defensible as such in court---and release the result under whatever terms they prefer.

This is not merely hypothetical. The most vivid public instance to date is the March~2026 \texttt{chardet} controversy: Dan Blanchard, long-time maintainer of the Python character-encoding library, used an AI coding agent to produce a ``ground-up, MIT-licensed rewrite'' of the previously LGPL-licensed codebase in roughly five working days, citing standard-library-inclusion requirements and a 48$\times$ speedup as motivation. A JPlag analysis reported under 1.3\% structural overlap with prior versions \cite{claburn2026chardet}.  The library's original author Mark Pilgrim, returning from more than a decade of public silence, objected that Blanchard's ``ample exposure'' to the LGPL codebase over a decade of maintenance made any clean-room defense implausible. Notably, Maffulli himself cites chardet as a paradigmatic illustration of the Second Liberation \cite{maffulli2026second}; read from the standpoint of this paper, the same event is a paradigmatic illustration of copyleft laundering. Chardet~7.0 represents the first high-profile case in which the engineering conditions for AI-assisted copyleft relicensing have become operational, whatever its eventual legal resolution.\footnote{At the time of writing, the matter has not entered the litigation record, and the legal question of whether chardet~7.0 is a derivative work of its LGPL predecessors remains unresolved. The claim here is that the engineering and economic conditions now permit such relicensing at a cost the copyleft tradition has never before had to confront---not that such laundering already succeeds as a legal matter.} The engineering workflow generalizes straightforwardly: feed the GPL source into the assistant, request a reimplementation in the same language or a different one, perhaps iterate a few times on architectural style, and ship. The resulting code is unlikely to contain substantial verbatim fragments of the original; under the idea/expression dichotomy, copyright protects expression and not the underlying ideas or functionality, leaving such rewrites with limited exposure \cite{samuelson2023genai}. This is not to say that every rewrite is safe: substantial similarity of non-literal elements can still support an infringement claim, and post-\emph{Google v.\ Oracle} doctrine remains unsettled on the boundary between functional and expressive reimplementation. But the threshold of safety is moving, and moving in favor of the evader. Even where a rewrite might qualify as a derivative work in the statutory sense, enforcement against a large vendor with deep pockets and plausible deniability would be prohibitively costly for most upstream maintainers. The cost of producing such rewrites is collapsing even faster than the cost of writing new code from scratch, because rewriting is a simpler task for current-generation models than greenfield development \cite{maffulli2026second}.

Copyleft was always dependent on a particular economic assumption: that the cost of rewriting a substantial codebase from scratch was high enough to make compliance cheaper than evasion. The GNU Compiler Collection, the Linux kernel, and readline---three canonical examples around which copyleft doctrine developed---are all systems whose reimplementation would historically have required years of skilled engineering labor. Generative AI dissolves this asymmetry. The threshold at which it becomes cheaper to evade copyleft than to comply with it is being pushed downward, library by library: small utility libraries are already well within the cost-effective evasion range, and only the very largest systems, whose complexity resists automated reconstruction, will retain the protection that cost asymmetry once afforded to all of free software. The process is gradual and largely invisible---there is no headline moment at which copyleft ``fails''---but its cumulative effect is to strip copyleft of the economic backbone on which its normative force depended.

The implication is direct. A governance regime for AGI-era software cannot rely on share-alike obligations over source code: a source file is no longer a scarce, artisanal artifact whose reproduction requires matching skill, but increasingly a cheap derivative of an expressible specification. What remains scarce---and therefore a plausible target for governance---is the verifiable procedure by which a system of a given behavioral profile came into existence. Crucially, the argument generalizes reflexively: if copyleft cannot meaningfully constrain AI-assisted rewriting of ordinary software, then \emph{a fortiori} it cannot constrain an AGI that rewrites its own training code, data-selection heuristics, or evaluation harness in pursuit of self-improvement. Requirement R6 (recursive verifiability), developed below, is our response to both at once.

\section{Reproducible Builds as a New Copyleft}
\label{sec:rb}

The free software community has been developing, for over a decade, an engineering practice that addresses exactly this kind of procedural audit: \emph{reproducible builds} \cite{reproduciblebuilds2024}. A build is reproducible if any party, starting from the same declared source and toolchain, obtains a bit-for-bit identical binary. The Debian, Tails, and NixOS projects, among others, have invested substantial effort in making their distributions reproducible; the motivation is that a disclosed source code cannot protect the user from a compromised build server unless the correspondence between source and binary can be independently checked \cite{lamb2022reproducible}.

Reproducible builds are thus the technical counterpart of the legal guarantee copyleft sought to provide. Where copyleft used the force of copyright to ensure that every binary was accompanied by its source, reproducible builds provide the cryptographic and procedural machinery to ensure that every binary is \emph{derivable from} its source. Together, the two practices discharge the source--object equivalence on which the four freedoms depend.

For AI systems, the question becomes: can a training pipeline, analogous to a compiler, be made reproducible? The honest answer is that full reproducibility of contemporary deep learning is difficult but not impossible, and that the technical obstacles are increasingly well understood.

\subsection{Sources of Non-Determinism in ML Training}

Chen et al.\ \cite{chen2022reproducible} offered a systematic taxonomy of reproducibility barriers for deep learning, dividing them into software randomness (PRNG seeds, data-loader shuffling, dropout masks) and hardware non-determinism (cuDNN algorithm selection, floating-point reduction order on GPUs). Their record-and-replay plus profile-and-patch framework was able to reproduce six open-source and one commercial deep-learning model exactly. Semmelrock et al.'s 2025 survey \cite{semmelrock2025barriers} expanded this analysis into a barriers-and-drivers matrix, identifying technology-driven, procedural, and educational levers and noting that reproducibility of large language models is particularly constrained by the sheer cost of retraining.

The U.S.\ Software Engineering Institute has argued that the non-determinism of ML is frequently overstated: with careful seed control and suppression of internal concurrency, ``the myth'' of ML irreproducibility collapses into a set of well-characterized engineering problems \cite{sei2025myth}. This is broadly consistent with the PyTorch project's own reproducibility guidance \cite{pytorch2024reproducibility}: with appropriate seed management and deterministic backend settings, training can be made repeatable on a given platform and release, though bit-exact reproducibility is generally not guaranteed across CPU/GPU boundaries, across GPU microarchitectures, or across framework versions.

\subsection{Deterministic Inference}

A recent and important development concerns inference rather than training. Thinking Machines Lab, in a widely discussed September~2025 blog post \cite{tml2025nondeterminism}, showed that the apparent non-determinism of temperature-zero LLM inference is not primarily due to floating-point non-associativity combined with GPU scheduling, but to the \emph{batch-size dependence} of reduction kernels: a given prompt can be dispatched under different dynamic batch sizes, and the internal reduction tree changes accordingly. The authors provide batch-invariant kernels for RMSNorm, matrix multiplication, and attention, and demonstrate bit-identical outputs across 1{,}000 repeated runs on Qwen3-8B, at approximately a 61.5\% throughput cost in their baseline implementation. The SGLang team \cite{lmsys2025deterministic} subsequently integrated these kernels with CUDA graphs, reducing the overhead to approximately 34.35\%. In collaboration with the slime project, they also extended the approach to fully reproducible reinforcement-learning training, closing the train--inference gap that had silently rendered on-policy RL off-policy. The LLM-42 project \cite{llm42} takes a different route: arguing that batch-invariant computation is fundamentally over-constrained because it strips GPU kernels of batch-adaptive parallelism strategies, it instead proposes a \emph{scheduling-based decode--verify--rollback} protocol inspired by speculative decoding, which enforces determinism selectively and incurs overhead roughly proportional to the fraction of traffic that actually requires it.

What these results demonstrate is that bit-exact reproducibility of large models is a \emph{tractable engineering problem} rather than a fundamental limit. The question is which parts of the pipeline need to be made deterministic, and at what cost, to satisfy a given governance requirement. This is the question the next section formulates as a requirements specification.

\section{Requirements for AGI-oriented Reproducible Builds}
\label{sec:requirements}

An AGI, as defined for the purposes of data governance in \cite{hatta2025agi} following Principle~22 of the Asilomar AI Principles \cite{asilomar2017}, is an AI system capable of recursive self-improvement or self-replication. Such a system poses distinctive reproducibility challenges that go beyond those identified for contemporary foundation models. We define a set of requirements for what we call an \emph{AGI-oriented reproducible build} (AGI-RB).

The seven requirements fall into three categories. R1--R5 are engineering requirements at varying levels of current maturity, applicable with increasing technical effort to existing and near-term systems. R6 is a \emph{research target} rather than a deployable specification: it identifies the correct invariant that a share-alike obligation must enforce in the AGI context, even though no existing architecture demonstrably satisfies it. R7 is a \emph{feasibility constraint} on the entire framework, governing the pace and scope at which the other requirements can realistically be imposed.

\subsection{R1. Complete Input Enumeration}

An AGI-RB must specify, in machine-readable form, the complete set of inputs whose combination determines the trained system. At a minimum this includes: the exact training corpus (not merely descriptive metadata); all preprocessing and tokenization code with pinned dependency versions; all hyperparameters; the initial weights or the seed used to generate them; the optimizer state at each checkpoint; and a hash-verifiable description of the toolchain and hardware configuration. OSAID's ``data information'' category is insufficient at this level: a substantially equivalent dataset will not yield bit-identical weights.

\subsection{R2. Deterministic Training Pipeline}

The training pipeline must be configured such that the declared inputs reproduce the declared outputs under bit-identity, either directly or after a known-equivalent transformation. Techniques include fixed PRNG seeds, deterministic cuDNN modes, batch-invariant reduction kernels \cite{tml2025nondeterminism}, and suppression of dynamic batching in distributed training. Where full determinism is infeasible for scale reasons, the pipeline must specify a \emph{reproducibility budget}: the admissible numerical tolerance and the statistical test used to certify it.

\subsection{R3. Verifiable Toolchain and Hardware Binding}

Because GPU microcode, CUDA versions, and interconnect topologies affect numerical results, an AGI-RB must either bind to specific hardware configurations (and disclose them) or ship with hardware-abstraction layers---patched libraries replacing non-deterministic primitives---as part of the distribution \cite{chen2022reproducible}. The practice parallels the \texttt{dpkg-buildflags} and cross-compilation hygiene established by the reproducible-builds community for ordinary software \cite{reproduciblebuilds2024}.

\subsection{R4. Third-Party Attestation Infrastructure}

Few end users will themselves retrain an AGI. As in software, where reproducible-builds testers such as \texttt{reproducible.debian.net} publish independent verification, AGI-RB requires a third-party attestation infrastructure. Hugging Face, MLCommons, or a consortium analogous to the Reproducible Builds project could operate public verification pipelines that retrain, or at minimum re-infer, claimed open source AI systems and publish cryptographic hashes of the results. Small language models, as noted in \cite{hatta2024copyleft}, are particularly amenable to this since the retraining cost falls within ordinary research budgets. Regulatory frameworks are moving in this direction: the EU AI Act's provisions for open-source foundation models \cite{eu2024aiact} already treat openness as an exemption criterion, and reproducibility is increasingly discussed as a natural operational test for that criterion \cite{openfuture2024aia}.

\subsection{R5. Self-Improvement Trajectory Logging}

This is the requirement distinctive to AGI. A system capable of recursive self-improvement modifies, at each iteration, both itself and the way in which it processes data \cite{yampolskiy2015}. A reproducible build of such a system must therefore include not only the initial training artifacts but the full trajectory of self-modification: every intermediate checkpoint, every modification to the training code generated by the system itself, and every dataset generated or curated by the system for its own use. In practice this implies an append-only, cryptographically signed log of self-modification events, analogous in spirit to Certificate Transparency logs in the TLS ecosystem.

\subsection{R6. Recursive Verifiability (Research Target)}

A stronger requirement follows from R5: the verification property must itself be preserved under self-improvement. If a system modifies its own training code, the modified training code must also produce reproducible outputs, and the modification must be auditable. This is a non-trivial constraint on AGI architecture design: it restricts the space of legitimate self-improvement operations to those that preserve the reproducibility invariant. It is, we suggest, the AGI-era analogue of the viral propagation that copyleft achieved through its share-alike clause. Instead of propagating a licensing term, the invariant propagates a technical property.

We acknowledge that R6 is an open research problem rather than a deployable engineering specification. Preserving a reproducibility invariant across arbitrary self-modifications touches on long-standing questions about inner alignment, tiling agents, and reflective stability \cite{yudkowsky2013tiling}, and it is not obvious that any AGI architecture proposed to date satisfies the property. Our claim is therefore the weaker one that R6 identifies the correct \emph{target}: whatever substantive form share-alike takes in the AGI context, it must operate at the level of invariants on self-modification rather than at the level of license terms on static artifacts.

\subsection{R7. Sustainable Economic and Computational Model}

Finally, any copyleft-like requirement that is economically infeasible will be ignored in practice. Demanding full reproducibility of a trillion-parameter frontier model is, at 2026 compute prices, not a meaningful obligation to impose on most actors. A realistic AGI-RB regime must therefore (i) begin with domain-restricted small models \cite{hatta2024copyleft}; (ii) exploit batch-invariant kernels \cite{tml2025nondeterminism,lmsys2025deterministic} and selective-determinism techniques such as decode--verify--rollback \cite{llm42} to contain the verification tax; and (iii) tolerate verifiable-by-sampling regimes in which third parties rebuild only subsets of the training trajectory.

A potential objection deserves acknowledgment: reproducibility requirements, like any compliance burden, could entrench large incumbents who can afford the necessary infrastructure, producing the opposite of the democratizing effect intended---a dynamic Widder, West, and Whittaker \cite{widder2023open} document in the political economy of ``open'' AI. The response lies in R7 itself: calibrating requirements to model scale and domain, phasing obligations as tooling matures, and situating attestation in neutral consortium infrastructure imposes comparable relative burdens rather than absolute costs only incumbents can absorb. As with the GPL---whose obligations scaled with acts of distribution rather than organizational size---an analogous scaling principle should govern AGI-RB.

\section{MCP and the Dynamic Linking Layer: A Different Governance Problem}
\label{sec:mcp}

Our analysis so far has concerned the production of AI systems: the training pipeline, the weights, the self-improvement trajectory. But AI systems increasingly operate not in isolation but as nodes in a rapidly growing web of tool-using and tool-providing services. A class of runtime coupling standards has emerged to regulate these interactions: OpenAI's function-calling interface \cite{openai2023functions}, Google's Agent-to-Agent (A2A) protocol \cite{google2025a2a}, and various framework-level tool abstractions in libraries such as LangChain all share the same architectural role. For concreteness, we focus in what follows on Anthropic's \emph{Model Context Protocol} (MCP), introduced in late 2024 and now implemented across a broad range of AI products \cite{anthropic2024mcp}; MCP standardizes how models discover and invoke external tools, resources, and prompts at runtime. A model with MCP access can read a user's calendar, query a database, post to a repository, or call another model's inference endpoint, all within the same conversational turn. Our argument is not specific to MCP, however: it applies \emph{mutatis mutandis} to any standard that plays the same coupling role. Where we write ``MCP'' below, the reader may substitute any sufficiently general successor or competitor protocol.

MCP is not, in itself, a training-time artifact; it is a runtime coupling mechanism. In this respect it stands in the same relation to the AI system that \emph{dynamic linking} stands to an executable in classical software. And this analogy is more than rhetorical: it has direct, and largely overlooked, implications for the governance framework developed in this paper.

\subsection{Why Copyleft Never Reached the Linking Layer}

The history of the GNU project is in part a history of litigated edge cases over where the copyleft obligation stops. The distinction between a derivative work and a mere aggregation, between static and dynamic linking, between invocation across a process boundary and invocation within it---these questions were never cleanly resolved. The LGPL exists precisely because the FSF decided, for pragmatic reasons, not to push copyleft across the linking boundary for libraries deemed infrastructural \cite{stallman2002free}. GPLv3 introduced carefully negotiated language around ``System Libraries'' and ``Corresponding Source'' to keep the viral effect within tractable limits \cite{fsf2007gpl3}.

The underlying difficulty is that dynamic linking dissolves the artifact that copyright law is designed to regulate. When program $A$ loads library $B$ at runtime, there is no distributed ``work'' that combines both; there are two works that interact in a user's memory. Copyright's grip weakens as we move from the static object file to the running process, and it disappears entirely when interaction takes place across a network boundary---the Affero clause of the GNU Affero General Public License (AGPLv3) being an elaborate attempt, widely regarded as partially successful at best, to plug that specific hole.

MCP makes this problem central rather than peripheral. A GPT-class model calling a third-party MCP server to retrieve a document is not distributing anything to anyone; it is performing a runtime interaction. If the model and the server are both ``open source AI'' in the OSAID sense, well and good; but nothing in the current OSAID, MOF, or OpenMDW frameworks says anything about the \emph{edges} that connect them. The licensing regime is a set of rules for what each node must disclose about itself, and is silent on the protocols by which nodes interact.

\subsection{The Masnick Insight: Protocols, Not Platforms}

Masnick \cite{masnick2019protocols}, writing about social media content moderation, argued that many of the pathologies of the platform era derive from the bundling, within a single corporate gatekeeper, of functions that the open-protocol era kept architecturally separate. Extending Masnick's argument, we can identify three such functions in particular: hosting, routing, and user interface. Centralized platforms bundle these into a single governance surface, which is why Facebook's content-moderation decisions attract the attention they do: the platform is simultaneously the host, the router, and the primary interface. A protocol-based architecture, Masnick argues, would push moderation decisions to the edges---to interfaces, clients, filters---while keeping the protocol itself neutral and open. The historical precedent is email: SMTP as a protocol remains universal, while thousands of clients, servers, and filtering services compete on the user-facing experience.

The parallel to the AI tool-use layer is striking. MCP, if it succeeds, will be to AI assistants roughly what SMTP is to email: a lingua franca that decouples the production of capabilities from the consumption of capabilities. A well-governed MCP ecosystem would then exhibit many of the properties Masnick attributes to protocol-based social media: competition at the client (assistant) layer, innovation at the tool-provider layer, and governance distributed across edges rather than concentrated in a single gatekeeper. Conversely, a poorly governed MCP ecosystem---one in which a single vendor controls the registry, the reference implementations, and the authentication infrastructure---would exhibit exactly the platform pathologies that Masnick diagnoses for current social media.

The analogy has limits. Masnick's concern was end-user speech and content-moderation pluralism, whereas the governance problem at the AI tool-use layer centers on tool-provider competition, portability of user tool ecosystems, and avoidance of vendor lock-in at the orchestration layer. These are different first-order concerns, and the Masnick framework cannot simply be imported wholesale. What does transfer, however, is the structural insight that separating the three bundled functions---hosting, routing, and interface---produces governance properties that no centralized platform design can match. It is this structural claim, not the content-moderation application, that we apply to MCP.

This suggests that the governance of the AI linking layer is a fundamentally different problem from the governance of the training layer, and requires different tools. Copyleft and reproducible builds are appropriate for the production side, where the question is how a system comes into being. Protocol governance in the Masnick sense is appropriate for the linking side, where the question is how systems interact. Conflating the two, or attempting to extend a licensing framework designed for one into the other, is likely to produce the same unsatisfying results that extending copyright law to cover dynamic linking produced in the 1990s.

\subsection{Requirements for MCP-Layer Governance}

A full development of this point would require another paper, but we can indicate the direction. An MCP-style protocol designed for good governance would need, at minimum: (i) an open, standards-body-maintained specification that no single vendor can unilaterally alter---exactly the property Masnick identifies as the core of the protocol/platform distinction; (ii) portable, non-proprietary authentication so that a user or organization can switch between clients without losing access to their tool ecosystem---the email analogue being the portability of one's address book across providers; (iii) transparent logging of tool invocations in a form the user controls, so that audit of AI behavior is possible without depending on the goodwill of any single vendor; and (iv) explicit, machine-readable licensing and provenance metadata on tools themselves, so that the copyleft status of downstream artifacts produced by tool composition can be determined rather than assumed.

The last of these is particularly important in light of the rewrite problem discussed in Section~\ref{sec:rewrite}. If an AI system with tool access can generate a new artifact by composing outputs from multiple MCP-accessed sources, the licensing status of the result is an open question that current MCP implementations do not attempt to answer. Metadata requirements at the protocol level would begin to address this, not by re-imposing copyleft viral propagation (which would be both technically unenforceable and contrary to the Masnick protocol-neutrality principle), but by enabling downstream reasoning about provenance.

We note, finally, that MCP-like protocols exhibit the same structural vulnerability to incumbent capture that Masnick identifies for social-media platforms. MCP is currently maintained by Anthropic, a single commercial entity; long-term governance of the specification is unsettled. If MCP becomes the AI industry's SMTP, its stewardship should reside in a neutral body along the lines of the IETF, not in any single laboratory. This is an institutional design question likely to prove as consequential as the OSAID debate, if not more so.

\section{Connection to AGI Data Governance}
\label{sec:governance}

In \cite{hatta2025agi} we identified seven data-governance challenges specific to AGI: unpredictability of data collection, divergent optimization criteria, AGI-to-AGI data sharing, provenance tracking under recursive self-improvement, intellectual-property complications, cross-border fragmentation, and temporal mismatch between governance frameworks and system evolution. The AGI-RB framework addresses several directly. Most importantly, R5 and R6 together impose technical preconditions on provenance tracking under recursive self-improvement: a self-modifying system whose modifications preserve reproducibility is amenable to after-the-fact audit in a way that a black-box system is not. R4 offers a partial answer to the temporal challenge, since governance can rely on continuously updated verification infrastructure rather than ex ante certification that inevitably goes stale.

The intellectual-property challenge intersects with the OSAID data compromise in a way that deserves closer attention than we can give here. If an AGI generates new training data for self-improvement, a reproducibility requirement implies that those generated datasets must be disclosable, raising the question of ownership \cite{samuelson2023genai}. Our tentative view, consistent with \cite{hatta2024copyleft}, is that a copyleft-like obligation attached to AGI-generated artifacts---requiring release under terms compatible with further reproducibility---may be the cleanest resolution: a genuine share-alike obligation whose content is reproducibility rather than source availability. More concretely, AGI systems deployed in domains of public interest---government procurement, defense, safety-critical medicine---should be required to satisfy at least R1--R4, with R5--R6 phased in as technical capability matures, structurally analogous to what the GPL achieved for federally funded software but implemented through reproducibility rather than licensing alone.

\section{Conclusion}
\label{sec:conclusion}

Copyleft was a legal hack that worked because a technical fact---the deterministic relationship between source and object code---made the legal obligation substantively meaningful. That fact no longer holds for modern AI systems, and will hold even less for AGI. Maffulli is right that the software-freedom tradition cannot simply be rewritten in AI terms using the old legal toolkit, but wrong to suggest that the solution is already arriving in the form of more powerful AI assistants. What is needed is a technical reconstruction of the source--object equivalence, adapted to systems whose behavior emerges from code, data, weights, seeds, toolchains, hardware, and increasingly from their own prior self-modifications.

Reproducible builds, extended from the software context to the full training and inference pipeline, are the natural candidate. We have formulated seven requirements for AGI-oriented reproducible builds---five engineering requirements at varying maturity levels, one research target (R6), and one feasibility constraint (R7)---connected them to ongoing international work, and situated them within the broader landscape of AGI data governance. Two further phenomena---AI-assisted rewriting of licensed code and the rise of MCP-style dynamic coupling---extend the diagnosis: for the former, reproducibility provides the only remaining point of purchase once cost asymmetry erodes; for the latter, Masnick's protocol-governance template offers a structural alternative to platform pathologies. The framework is preliminary, and much remains to be worked out, but the direction is correct.

The first liberation, to borrow Maffulli's phrase, gave us the code. The second liberation, if it is to mean anything more than a new dependency on opaque AI services, must give us the code \emph{and} the verifiable procedure by which the code became what it is. For AGI, anything less is not freedom but faith.

\paragraph{Acknowledgements.} This work was supported by JSPS KAKENHI Grant Number 26K15531.

\paragraph{Disclosure of Interests.} The author has no competing interests to declare that are relevant to the content of this paper. The author is a member of the Open Source Initiative community and has participated in OSAID discussions.

\end{document}